\documentclass[letterpaper, 10 pt, conference]{ieeeconf}  

\IEEEoverridecommandlockouts                              
\usepackage{amsfonts}
\usepackage{graphicx,xcolor}
\usepackage{amsmath}
\usepackage{amssymb}
\usepackage{dsfont}

\usepackage{ifthen}
\usepackage{color}
\usepackage{cite}

\usepackage{mathtools}
\usepackage{booktabs}
\usepackage{url}
\usepackage{hyperref}
\usepackage{caption}
\usepackage{subfig}
\usepackage{float}
\usepackage{multirow}

\def\scr#1{{\cal #1}}
\newcommand{\R}{{\rm I\!R}}
\def\eq#1{\begin{equation}#1\end{equation}}
\def\rep#1{(\ref{#1})}
\newcommand{\bbb}{\mathbb}
\newtheorem{theorem}{Theorem}
\newtheorem{lemma}{Lemma}

\newtheorem{definition}{Definition}
\newtheorem{proposition}{Proposition}
\newtheorem{corollary}{Corollary}

\newcommand{\dfb}{\stackrel{\Delta}{=}}
\def\qed{ \rule{.08in}{.08in}}

\newcommand{\0}{\mathbf{0}}

\DeclareMathOperator*{\argmin}{arg\,min}

\title{\LARGE \bf A Resilient Distributed Algorithm for Solving Linear Equations
}

\author{Jingxuan Zhu \hspace{.3in} Alvaro Velasquez \hspace{.3in} Ji Liu
\thanks{
J. Zhu is with the Department of Applied Mathematics and Statistics at Stony Brook University (\texttt{jingxuan.zhu@stonybrook.edu}).
A.~Velasquez is with the Department of Computer Science at University of Colorado Boulder (\texttt{alvaro.velasquez@colorado.edu}).
J. Liu is with the Department of Electrical and Computer Engineering at Stony Brook University
(\texttt{ji.liu@stonybrook.edu}).
}
}

\begin{document}

\maketitle
\thispagestyle{empty}
\pagestyle{empty}


\begin{abstract}
This paper presents a resilient distributed algorithm for solving a system of linear algebraic equations over a multi-agent network in the presence of Byzantine agents capable of arbitrarily introducing untrustworthy information in communication. It is shown that the algorithm causes all non-Byzantine agents' states to converge to the same least squares solution exponentially fast, provided appropriate levels of graph redundancy and objective redundancy are established. An explicit convergence rate is also provided. 
\end{abstract}


\section{Problem}

There has been considerable interest in designing distributed algorithms for solving a possibly large system of linear algebraic equations over a multi-agent network, stemming from the work of \cite{le}. A review of this topic can be found in \cite{review}. While various problem formulations have been proposed and studied, following \cite{le} we focus on the following basic and important information distribution~setting. 

Consider a multi-agent network consisting of $n$ agents labeled $1$ through $n$ for the purpose of presentation. Each agent is not aware of such a global identification number, but is capable of distinguishing between its neighbors. The neighbor relations among the $n$ agents are characterized by a directed graph $\bbb{G} = (\mathcal{V},\mathcal{E})$ whose
vertices correspond to agents and whose directed edges (or arcs) depict neighbor relations, where $\mathcal{V}=\{1,\ldots,n\}$ is the vertex set and $\mathcal{E}\subset\mathcal{V} \times \mathcal{V}$ is the directed edge set.\footnote{We use $\scr{A} \subset \scr{B}$ to denote that $\scr A$ is a subset of $\scr B$.}
We say that agent~$i$ is a neighbor of agent $j$ if $(i,j)\in\scr{E}$. The directions of arcs represent the directions of information flow in that each agent can receive information only from its neighbors. 

Each agent $i\in\scr V$ knows a pair of ``private'' real-valued  matrices $(A_i^{r_i\times d},b_i^{r_i\times 1})$ which are only known to agent $i$. The problem of interest is to devise local algorithms,
 one for each agent, which will enable all $n$ agents to simultaneously and iteratively compute
   the same least squares solution
 to the linear algebraic equation $Ax=b$
where
$$A = \begin{bmatrix}A_1 \cr A_2 \cr \vdots \cr A_n\end{bmatrix}_{r\times d}  \;\;\;\; {\rm and} \;\;\;\;\;\;
b= \begin{bmatrix}b_1\cr b_2\cr \vdots \cr b_n\end{bmatrix}_{r\times 1}$$
with  $r =\sum_{i=1}^n r_i$.
Such a distributed least squares problem has been studied in the literature, e.g., \cite{le,mou_ls,guodong,meng,hyosung}.
Let 
$$\scr X^* \dfb \argmin_x \|Ax-b\|_2^2 = \argmin_x\sum_{i=1}^n \|A_ix-b_i\|_2^2$$ 
be the set of all least squares solutions, where $\|\cdot\|_2$ denotes the 2-norm. It is easy to see from the above equality that the distributed least squares problem can be reformulated as a distributed convex optimization problem and thus is solvable via a vast number of existing distributed optimization algorithms \cite{nedic2018distributed,yang2019survey}. 
It is well known that $\scr X^* = \{x:A'Ax=A'b\}$ and is always nonempty. We
will use this fact of least squares solutions without special mention in the sequel.

In this paper we consider a more challenging variant of the distributed least squares problem in the presence of Byzantine agents capable of transmitting arbitrary values to other agents and transferring conflicting values to different agents at any time. We use $\scr F$ to denote the set of Byzantine agents and $\scr H$ to denote the set of normal (non-Byzantine) agents. Which agents are Byzantine is unknown to normal agents. 
It is assumed that each agent may have at most $\beta$ Byzantine neighbors. The goal of the normal agents is to cooperatively reach a consensus at the same least squares solution to $Ax=b$. 

From the preceding discussion, the resilient distributed least squares problem just described can also be treated as a resilient distributed convex optimization problem. It turns out that very few papers accurately solve this resilient problem with theoretical guarantees. 


The resilient distributed optimization algorithms in \cite{acc2023,toread} are expected to be applicable to the problem under consideration with some modification. The algorithm in \cite{acc2023} is based on the subgradient method (for convex but not necessarily differentiable objective functions) and a nonempty $\scr X^*$ interior assumption (cf. Assumption 1 in \cite{acc2023}), and thus specialization of that algorithm to the problem of interest here needs a more careful treatment. Correctness of the algorithm in \cite{toread} is built upon strong convexity of the objective function, so it can be used to solve the resilient least squares problem only when $Ax=b$ has a unique least squares solution, which narrows scope of the problem.  
Moreover, both algorithms in \cite{acc2023} and \cite{toread} make use of time-varying diminishing stepsizes and thus cannot converge exponentially fast. It is worth mentioning that the state-of-the-art least squares and distributed least squares algorithms achieve (at least) exponential convergence. 
It will be clear shortly that the problem under consideration is closely related to a resilient version of so-called constrained consensus problem \cite{constrain}. In general a discrete-time constrained consensus cannot be reached exponentially fast unless a certain constrained set regularity condition is satisfied and exploited in analysis \cite{tacrate}. We will further comment on this point in the next section.

This paper proposes a resilient distributed least squares algorithm with guaranteed exponentially fast convergence. The algorithm follows two quantifiable redundancy notions in \cite{acc2023},\footnote{The two notions were prompted by preexisting ones, e.g., \cite{reviewer} and \cite{Va12}; see detailed discussion in \cite[Sections 1 and 2.1]{acc2023}.} 
namely objective redundancy and graph redundancy, with the former being tailored for distributed linear equations. 
An explicit rate of convergence is derived, reflecting the effects of quantified levels of both graph redundancy and objective redundancy.

\section{Redundancy}

It is easy to see that a multi-agent system without an attack detection/correction capability is unable to solve the resilient distributed least squares problem unless certain redundancy is established. We begin with the objective redundancy.

\vspace{.05in}

\begin{definition}\label{def:least_squares_redundant}
An $n$-agent network is called $k$-redundant, $k\in\{0,1,\ldots,n-1\}$, if for any subsets $\scr S_1,\scr S_2\subset\scr V$ with $|\scr S_1|= |\scr S_2| = n - k$, there holds\footnote{We use $|\scr S|$ to denote the cardinality of a set $\scr S$.} 
\begin{align*}
    \argmin_{x}\sum_{i\in\scr S_1}
    \|A_ix-b_i\|_2^2=\argmin_{x}\sum_{i\in\scr S_2}\|A_ix-b_i\|_2^2.
\end{align*} 
\end{definition}

\vspace{.1in}
Note that for any nonempty agent subset $\scr U\subset\scr V$, $\argmin_{x}\sum_{i\in\scr U} \|A_ix-b_i\|_2^2$ is the least squares solution
 to the linear equation $A_{\scr U}x=b_{\scr U}$
where
$$A_{\scr U} = \begin{bmatrix}A_{\pi(1)} \cr A_{\pi(2)} \cr \vdots \cr A_{\pi(|\scr U|)}\end{bmatrix},  \;\;\;\;\;\;
b= \begin{bmatrix}b_{\pi(1)}\cr b_{\pi(1)}\cr \vdots \cr b_{\pi(|\scr U|)}\end{bmatrix},$$
and $\pi:\scr U\rightarrow\scr U$ is any permutation map. Thus, $$\argmin_{x}\sum_{i\in\scr U} \|A_ix-b_i\|_2^2 = \{x:A'_{\scr U}A_{\scr U}x=A'_{\scr U}b_{\scr U}\}.$$
The above objective redundancy is a well-defined quantifiable notion because of the following properties.   
\vspace{.05in}

\begin{lemma}\label{lem:equal_global_least_squares}
If an $n$-agent network is $k$-redundant, then for any subsets $\scr S,\scr L\subset\scr V$ with $|\scr S|= n-k$ and $|\scr L|\ge n-k$, 
\[\argmin_{x}\sum_{i\in\scr S}
    \|A_ix-b_i\|_2^2=\argmin_{x}\sum_{i\in\scr L}\|A_ix-b_i\|_2^2.\]
\end{lemma}

\vspace{.1in}

The lemma is essentially a special case of Lemma 1 in \cite{acc2023} 
and immediately implies the following results.

\vspace{.05in}

\begin{corollary}\label{coro:global_least_squares}
If an $n$-agent network is $k$-redundant, then for any subset $\scr S\subset\scr V$ with $|\scr S|\ge n-k$, there holds
\[\argmin_{x}\sum_{i\in\scr S}
    \|A_ix-b_i\|_2^2=\scr X^*.\]
\end{corollary}

\vspace{.1in}

\begin{corollary}
If an $n$-agent network is $(k+1)$-redundant with $k\ge 0$, then it is $k$-redundant.
\end{corollary}

\vspace{.05in}

More can be said. 


\begin{proposition}\label{prop:common_least_squares}
    If an $n$-agent network is $k$-redundant with $k\ge 1$, then 
$\scr X^*= \bigcap_{i=1}^n\{x: A'_iA_ix = A'_ib_i\}$.
\end{proposition}

\vspace{.05in}

{\bf Proof of Proposition \ref{prop:common_least_squares}:}
Note that $A'A = \sum_{i=1}^nA'_iA_i$ and $A'b = \sum_{i=1}^nA'_ib_i$. Then, $\scr X^*=\{x:\sum_{i=1}^nA_i'A_ix =\sum_{i=1}^nA_i'b_i\}$, which implies that $\scr X^*\supset\bigcap_{i=1}^n\{x: A'_iA_ix = A'_ib_i\}$. To prove the lemma, it suffices to prove $\scr X^*\subset\bigcap_{i=1}^n\{x: A'_iA_ix = A'_ib_i\}$, or equivalently, $\scr X^*\subset\{x: A'_iA_ix = A'_ib_i\}$ for all $i\in\scr V$. To this end, pick any $j\in\scr V$ and $x^*\in\scr X^*$. From Corollary \ref{coro:global_least_squares} with $k\ge 1$, $$x^*\in \argmin_{x}\sum_{i\in\scr V\setminus\{j\}}
    \|A_ix-b_i\|_2^2,$$
which implies that $\sum_{i\in\scr V\setminus\{j\}}A'_iA_ix^*=\sum_{i\in\scr V\setminus\{j\}}A'_ib_i$. The difference between this equality and $\sum_{i=1}^nA_i'A_ix^* =\sum_{i=1}^nA_i'b_i$ yields $A'_jA_jx^*=A'_jb_j$.
\hfill$\qed$

\vspace{.05in}

Proposition \ref{prop:common_least_squares} implies that with objective redundancy all agents share at least one common least squares solution and thus the resilient distributed least squares problem boils down to a resilient distributed linear equation problem. The algorithm to be presented exploits this important fact. 

Proposition \ref{prop:common_least_squares} also maps the resilient problem under study to a resilient constrained consensus problem. Resilient constrained consensus has been partially solved in \cite{acc22} only for complete graphs and studied in \cite{resilientconstrained} with an incomplete proof. It is worth emphasizing that discrete-time constrained consensus, first proposed in \cite{constrain}, in general does not enjoy exponentially fast convergence (see \cite[Proposition~2]{constrain}, \cite[Theorem 6]{tacrate}, and \cite[Theorem 1]{ren_constrained}), let alone a resilient variant. An exponentially fast constrained consensus can be achieved when all agents' local constrained sets meet a so-called set regularity condition, subsuming linear (in)equalities as a special case, but its analysis requires a more careful treatment even for a non-Byzantine multi-agent network \cite[Section V.C]{tacrate}.

We will specialize a constrained consensus approach, combining a multi-dimensional resilient consensus idea in \cite{acc2023}, to the resilient problem of interest here and appeal to a distributed linear equation analysis tool developed in \cite{le,le_automatica}. This combination yields a fully resilient distributed least squares algorithm with any arbitrary initialization and an exponential convergence guarantee. Although exponential convergence is an intuitive result at the first glance, its establishment is not trivial. This is because major existing analyses for non-Byzantine distributed linear equations and constrained consensus processes over time-varying graphs are based on jointly strongly connected graphs \cite{le} or jointly strongly connected components with the same vertex subset\footnote{A lifting approach is typically used to analyze these discrete-time distributed algorithms with bounded delays in which a nominal time-dependent expanded graph replaces the underlying neighbor graph. Jointly strongly connected neighbor graphs are jointly strongly connected components of the nominal expanded graph which thus share the same vertex subset.}  \cite{ren_constrained,tac_le_asyn},
whereas the analysis for the resilient algorithm here will cope with time-varying rooted graphs whose roots (maximal strongly connected components) may arbitrarily change over time (cf. Lemma \ref{lem:root}). The definitions of rooted and strongly connected graphs are given as follows.

Let us call a vertex $i$ in a directed graph $\bbb G$ a root of $\bbb G$ if for each other vertex $j$ of $\bbb G$, there is a directed path from $i$ to $j$. In other words, $i$ is a root of $\bbb G$ if it is the root of a directed spanning tree of $\bbb G$. We say that $\bbb G$ is rooted at $i$ if $i$ is in fact a root. 
It is not hard to see that a rooted graph $\bbb G$ has a unique maximal strongly connected component whose vertices are all roots of $\bbb G$.
A directed graph $\bbb G$ is called strongly connected if there is a directed path from every vertex to every other vertex. Thus, every vertex in a strongly connected graph is a root. Any strongly connected graph must be rooted, but not vice versa.

We will also need a graph redundancy notion from \cite{acc2023}. 

\vspace{.05in}

\begin{definition}\label{def:resilient}
An $(r,s)$-reduced graph of a directed graph $\bbb G$ with $n$ vertices, with $r,s\ge 0$ and $r+s\le n-1$, is a subgraph of $\bbb G$ obtained by first picking any vertex subset $\scr S\subset\scr V$ with $|\scr S|=n-r$ and then removing from each vertex of the subgraph induced by $\scr S$, $\bbb G_{\scr S}$, arbitrary $s$ incoming edges in $\bbb G_{\scr S}$. 
A directed graph $\bbb G$ is called $(r,s)$-resilient if all its $(r,s)$-reduced graphs are rooted.
\end{definition}

\vspace{.05in}

It is easy to see that if a directed graph is $(r_1,s_1)$-resilient, then for any nonnegative $r_2\le r_1$ and $s_2\le s_1$, the graph is also $(r_2,s_2)$-resilient.
More can be said. If a directed graph is $(r,s)$-resilient, each of its vertices has at least $(r+s+1)$ neighbors \cite[Lemma 2]{acc2023}.

%
%

Equipping a multi-agent network with certain levels of both objective and graph redundancy, allows us to present the following feasible resilient algorithm.

\section{Algorithm}

Each agent $i$ has a time-dependent
state vector $x_i(t)$ taking values in $\R^d$, which represents its estimate of a least squares solution at time $t$ and can be arbitrarily initialized at $t=0$. 
It is assumed that the information agent
 $i$ receives from a normal neighbor $j$ is only the current state vector of neighbor~$j$. The algorithm will make use of the multi-dimensional resilient consensus idea\footnote{The idea was prompted by and simplified from the preexisting one in \cite{resilientconstrained}; see detailed discussion in \cite[Sections 1]{acc2023}.} 
 in \cite{acc2023} which needs the following notation. 

Let $\scr N_i$ be the set of neighbors of agent $i$ in the neighbor graph $\bbb G$ and $\scr A_i$ denote the collection of all those subsets of $\scr N_i$ whose cardinality is $(d+1)\beta + 1$. It is easy to see that the number of all such subsets is
\eq{a_i \dfb \binom{|\scr N_i|}{(d+1)\beta + 1},\label{eq:a_i(t)}}
assuming $|\scr N_i|\ge (d+1)\beta + 1$, and label them $\scr A_{i1},\ldots, \scr A_{ia_i}$.
For each $j\in\{1,\ldots,a_i\}$, let $\scr B_{ij}$ denote the collection of all those subsets of $\scr A_{ij}$ whose cardinality is $d\beta + 1$.
For any agent $i$ and any subset of its neighbors $\scr S\subset \scr N_i$, we use $\scr C_{i\scr S}(t)$ to denote the convex hull of all $x_{ji}(t)$, $j\in\scr S$ where $x_{ji}(t)$ denotes the vector agent $j$ sends to agent $i$. If agent $j$ is a normal agent, $x_{ji}(t)=x_j(t)$ for all possible $i$. If agent $j$ is a Byzantine agent, each $x_{ji}(t)$ can be an arbitrary vector. 



{\bf Algorithm:} At each discrete time $t\in\{0,1,2,\ldots\}$, each agent $i$ first picks an arbitrary point 
\eq{y_{ij}(t)\in\bigcap_{\scr S\in\scr B_{ij}} \scr C_{i\scr S}(t)\label{eq:hull}}
for each $j\in\{1,\ldots,a_i\}$, and 
then updates its state by setting
\begin{align}
    v_i(t) &= \frac{1}{1+a_i}\Big(x_i(t)+\sum_{j=1}^{a_i}y_{ij}(t)\Big),\label{eq:v_i(t)_ori}\\
    x_i(t+1) &= P_iv_i(t) + (A'_iA_i)^{\dagger}A'_ib_i,\label{eq:x_ori_ls}
\end{align}
where $P_i$ is the orthogonal projection on the kernel of $A_i$ and $M^{\dagger}$ denotes the Moore-Penrose inverse of matrix $M$. 
\hfill$\Box$


\vspace{.05in}

Update \eqref{eq:v_i(t)_ori} is the multi-dimensional resilient consensus step. The idea behind update \eqref{eq:x_ori_ls} is as follows. Note that
\eq{P_i=I-A_i^{\dagger}A_i = I - (A'_iA_i)^{\dagger}A'_iA_i\label{eq:samekernel}}
since the kernel of $A'_iA_i$ equals the kernel of $A_i$.
Using the standard quadratic programming with equality constraints, it is straightforward to show that $x_i(t+1)$ in \eqref{eq:x_ori_ls} is a solution to $\min_x\|x-v_i(t)\|_2$ subject to $A'_iA_ix=A_ib_i$. From this point of view, updates \eqref{eq:v_i(t)_ori}--\eqref{eq:x_ori_ls} can be regarded as a resilient variant of affine equality constrained consensus \cite{tacrate}. 

To state the main result, we need the following notation. For a directed graph $\bbb G$, let $\scr R_{r,s}(\bbb G)$ denote the set of all $(r,s)$-reduced graphs of $\bbb G$. For a rooted graph $\bbb G$, we use $\kappa(\bbb G)$ to denote the size of the unique maximal strongly connected component whose vertices are all roots of $\bbb G$; in other words, $\kappa(\bbb G)$ equals the number of roots of $\bbb G$. For any $(r,s)$-resilient graph $\bbb G$, define 
$$\kappa_{r,s}(\bbb G)\dfb\min_{\bbb H\in\scr R_{r,s}(\bbb G)} \kappa(\bbb H),$$
which denotes the smallest possible number of roots in any $(r,s)$-reduced graphs of $\bbb G$.


\vspace{.05in}

\begin{theorem}\label{thm:le}
If $\bbb G$ is $(\beta,d\beta)$-resilient and the $n$-agent network is $(n-\kappa_{\beta,d\beta}(\bbb G))$-redundant, then all $x_i(t)$, $i\in\scr H$ will converge to the same least squares solution to $Ax=b$ exponentially fast. 
\end{theorem}

\vspace{.05in}

A $(\beta,d\beta)$-resilient $\bbb G$ ensures that each agent has at least $(d+1)\beta + 1$ neighbors at each time $t$ \cite[Lemma 2]{acc2023}. 
This further guarantees that $y_{ij}(t)$ in \eqref{eq:hull} must exist \cite[Lemma~5]{acc2023}.


Theorem \ref{thm:le} is a direct consequence of Theorems \ref{main1} and \ref{main2} in the next subsection. 
From the proof of Theorem \ref{main1} and using the same arguments as in the proof of Corollary 1 in \cite{le}, it is straightforward to obtain the following convergence rate result of the algorithm.

\vspace{.05in}

\begin{corollary}
Suppose that $Ax = b$ has a unique least squares solution. If $\bbb G$ is $(\beta,d\beta)$-resilient and the $n$-agent network is $(n-\kappa_{\beta,d\beta}(\bbb G))$-redundant, then all $x_i(t)$, $i\in\scr H$ converge to the least squares solution as $t\rightarrow\infty$
at the rate as $\lambda^t$ converges to zero, where 
$$\lambda = \Big(1-(|\scr H|-1)(1-\rho)\eta^\tau\Big)^{\frac{1}{\tau}},$$
$\tau$
is a positive integer such that
for any  $p\geq \tau$,  the matrix
$P(W(t+p)\otimes I)\cdots  P(W(t+1)\otimes I)P(W(t)\otimes I)$
  is a contraction in the mixed  matrix norm for all $t\ge 0$, $\rho = \max_{\scr C} \|P_{j_1}P_{j_2}\cdots P_{j_{\tau+1}}\|_2$ with $\scr C$ being the set of all those products of the orthogonal projection  matrices in $\{P_1,P_2,\ldots,P_{|\scr H|}\}$ of length $\tau+1$
which are complete, and $\eta$ is defined in \eqref{eq:eta}.
\end{corollary}

\vspace{.05in}

The definition of ``complete'' products of orthogonal projection matrices will be given in the next subsection. It guarantees that, with set $\scr{C}$ being compact, $\rho$ is strictly less than one. With this fact, it is not hard to show that $\lambda\in [0,1)$  provided more than one normal agent exists in the network. It is worth mentioning that, from the proof of Proposition~\ref{p}, the value of $\tau$ is influenced by the levels of both graph redundancy and objective redundancy. 

The above convergence rate result can be straightforwardly extended to the case when $Ax = b$ has more than one least squares solution using the proof of Theorem \ref{main2}.


\subsection{Analysis}




To analyze the algorithm, we first derive the dynamics of normal agents which reject the influence of Byzantine agents because of step \eqref{eq:hull} of the algorithm. To this end, we need the following lemma. 

\vspace{.05in}

\begin{lemma}\label{le:convexcombforhighd}
{\rm \cite[Lemma 6]{acc2023}}
$v_i(t)$ in \eqref{eq:v_i(t)_ori} can be expressed as a convex combination of $x_i(t)$ and $x_k(t)$, $k\in\scr N_i\cap\scr H$, 
\begin{align}\label{eq:convexcombhighd}
    v_i(t) = w_{ii}(t)x_i(t) + \sum_{k\in\scr N_i\cap\scr H}w_{ik}(t)x_k(t),
\end{align}
where $w_{ii}(t)$ and $w_{ik}(t)$ are nonnegative numbers satisfying $w_{ii}(t)+ \sum_{k\in\scr N_i\cap\scr H}w_{ik}(t)=1$, and there exists a positive constant $\eta$ such that for all $i\in\scr H$ and $t$, $w_{ii}(t)\ge \eta$ and among all $w_{ik}(t)$, $k\in\scr N_i\cap\scr H$, at least 
$|\scr N_i\cap\scr H|-d\beta$ of them are bounded below by $\eta$.
\end{lemma}


\vspace{.05in}

From \eqref{eq:x_ori_ls} and Lemma \ref{le:convexcombforhighd}, the updates of all normal agents can be written as 
\begin{align}
    x_i(t+1) =\;& P_i\Big(w_{ii}(t)x_i(t) + \sum_{k\in\scr N_i\cap\scr H}w_{ik}(t)x_k(t)\Big) \nonumber\\
    &+ (A'_iA_i)^{\dagger}A'_ib_i,\;\;\;\;\; i\in\scr H,\label{eq:x_ls_ana}
\end{align}
which decouples from the dynamics of Byzantine agents.
Let $x^*$ be an arbitrary point in $\scr X^*$ and define $y_i(t)=x_i(t)-x^*$ for any $i\in\scr H$. Then, from \eqref{eq:x_ls_ana}, for all $i\in\scr H$,
\begin{align}
    y_i(t+1) =\;& P_i\Big(w_{ii}(t)y_i(t) + \sum_{k\in\scr N_i\cap\scr H}w_{ik}(t)y_k(t)\Big) \nonumber\\
    &+ (A'_iA_i)^{\dagger}A'_ib_i + P_ix^*-x^* \nonumber\\
    =\;& P_i\Big(w_{ii}(t)y_i(t) + \sum_{k\in\scr N_i\cap\scr H}w_{ik}(t)y_k(t)\Big), \label{eq:y_i_update}
\end{align}
where we used the fact that $(A'_iA_i)^{\dagger}A'_ib_i + P_ix^*-x^*=0$ for all $i\in\scr H$ which can be straightforwardly proved by \cite[Theorem 2]{mpinverse} and \eqref{eq:samekernel}.


Without loss of generality, we label all normal agents from $1$ to $|\scr H|$, i.e., $\scr H=\{1,2,\ldots,|\scr H|\}$, in the sequel. 

To proceed, let $y(t)$ denote a stack of all $y_i(t)$, $i\in\scr H$ with the
index in a top-down ascending order, i.e., 
$y(t)= [y'_1(t),y'_2(t),\ldots,y'_{|\scr H|}(t)]'$.
Then, the updates in \eqref{eq:y_i_update} can be written in the form of a state equation:
\begin{align}
    y(t+1) = P\big(W(t)\otimes I\big)y(t), \label{eq:normalxstate_ls}
\end{align}
where each $W(t)=[w_{ij}(t)]$ is a $|\scr H|\times |\scr H|$ stochastic matrix with positive diagonal entries\footnote{A square nonnegative matrix is called a stochastic matrix if its row sums all equal one.}, $\otimes$ denotes the Kronecker product, $I$ denotes the $d\times d$ identity matrix, and $P$ is a $d|\scr H|\times d|\scr H|$ block diagonal matrix whose $i$th diagonal block is $P_i$, $i\in\scr H$. It is easy to see that $P$ is also an orthogonal projection matrix.

Define the graph of an $m\times m$ matrix $M$ as a direct graph with $m$ vertices and an arc from vertex $i$ to vertex $j$ whenever the $ji$-th entry of $M$ is nonzero.
We will write $\gamma(M)$ for the graph of a matrix $M$.

\vspace{.05in}

\begin{lemma}\label{lem:root}
    {\rm \cite[Lemma 7]{acc2023}}
    If $\bbb G$ is $(\beta,d\beta)$-resilient, the graph of each $W(t)$ in \eqref{eq:normalxstate_ls} has a rooted spanning subgraph and all the diagonal entries and those off-diagonal entries of $W(t)$ corresponding to the rooted spanning subgraph are uniformly bounded below by a positive number 
    \eq{\eta\;\dfb\; \min_{i\in\scr V}\;\frac{1}{(d\beta+1)(1+a_i)\binom{(d+1)\beta+1}{d\beta+1}}.\label{eq:eta}}
\end{lemma}

\vspace{.1in}


It is worth emphasizing that although \eqref{eq:normalxstate_ls} shares almost the same state equation as that in \cite{le}, which is $y(t+1) = P(S(t)\otimes I)Py(t)$, a critical difference between the two is that the graphs of stochastic matrices $S(t)$ in \cite{le} are (jointly) strongly connected, whereas each graph of stochastic matrix $W(t)$ here is rooted (cf. Lemma \ref{lem:root}) with possibly  different roots. This is why analysis of the algorithm needs more careful treatment.

We first consider the case when $Ax=b$ has a unique least squares solution, i.e., $x^*\in\scr X^*$ is unique. From Proposition~\ref{prop:common_least_squares}, all $A_ix=b_i$, $i\in\scr V$ share a unique least squares solution. Let $\scr{P}_i$ denote the column span of $P_i$ for all $i$. 
Since $\scr P_i = {\rm kernel}\; A_i ={\rm kernel}\; A'_iA_i$, the least squares solution being unique is equivalent to 
\eq{\bigcap_{i=1}^n \scr{P}_i = 0.\label{assmp}}
More can be said. The following lemma is a direct consequence of Corollary \ref{coro:global_least_squares}. 

\vspace{.05in}

\begin{lemma}\label{lem:roots_uniqueness}
    If the $n$-agent network is $k$-redundant and \eqref{assmp} holds, then for any subset $\scr S\subset\scr V$ with $|\scr S|\ge n-k$, there holds $\bigcap_{i\in\scr S} \scr{P}_i = 0$.
\end{lemma}

\vspace{.05in}




We appeal to some concepts and results from \cite{reachingp1,le_automatica}.
Let us agree to call a vertex $i$ in a directed graph $\mathbb{G}$  a sink  of $\mathbb{G}$
  if for any other vertex $j$  of $\mathbb{G}$, there is a directed  path from vertex $j$ to vertex $i$.
 We say that
$\mathbb{G}$ is sunk  at $i$  if $i$ is in fact  a sink, 
and that $\mathbb{G}$ is  strongly sunk at $i$ if $i$ is  reachable from  each other vertex of $\mathbb{G}$
  along a  directed path of length one, i.e., any other vertex is a neighbor of vertex~$i$. 
A directed graph is called a sunk graph if  
it possesses at least one sink, 
and a strongly sunk graph if it has at least  one vertex at which it is strongly sunk. 

The composition of two directed graphs $\mathbb{G}_p$, $\mathbb{G}_q$
with the same vertex set,   denoted by $\mathbb{G}_q\circ\mathbb{G}_p$,
is the directed graph with the same vertex set and
 arc set defined  so that $(i, j)$ is an arc in the
composition whenever there is a vertex $k$ such that $(i, k)$
 is an arc in $\mathbb{G}_p$ and $(k, j)$
is an arc in $\mathbb{G}_q$.  Since this composition is an associative binary operation, the definition extends unambiguously to any finite sequence of directed graphs with the same vertex set.
Graph composition and matrix multiplication are closely related in that  
   $\gamma(M_2M_1) = \gamma(M_2)\circ\gamma(M_1)$.
For graphs with self-arcs at all
vertices, it is easy to see  that the arcs of both $\bbb{G}_p$ and $\bbb{G}_q$ are
arcs of $\bbb{G}_q \circ \bbb{G}_p$. 

\vspace{.05in}

\begin{lemma}
{\rm \cite[Lemma 5]{le_automatica}}
Let $\bbb{G}_{p_1},\bbb{G}_{p_2},\ldots,\bbb{G}_{p_k}$ be a finite sequence of $m$-vertex
directed graphs with self-arcs which are all sunk at $v$.
If $k \leq m-1$,
then $v$ has at least $k+1$ neighbors in $\bbb{G}_{p_k}\circ\bbb{G}_{p_{k-1}}\circ\cdots\circ\bbb{G}_{p_1}$.
If $k \ge m-1$,
then $\bbb{G}_{p_k}\circ\bbb{G}_{p_{k-1}}\circ\cdots\circ\bbb{G}_{p_1}$ is strongly sunk at $v$.
\label{kobs}\end{lemma}

\vspace{.05in}

Define the neighbor function of a directed graph $\bbb{G}$ with vertex set $\scr{V}$,
denoted by $\alpha(\bbb{G},\cdot)$,
as the $2^{\scr{V}}\rightarrow 2^{\scr{V}}$ function which assigns to each subset $\scr{S}\subset\scr{V}$, the subset of vertices in $\scr{V}$
which are neighbors of $\scr{S}$ in $\bbb{G}$.
For any $v\in\scr{V}$, there is a unique largest subgraph sunk at $v$, namely
the graph induced by the vertex set
$\scr{V}(v) = \{v\}\cup\alpha(\bbb{G},v)\cup \cdots \cup \alpha^{|\scr V|-1}(\bbb{G},v)$,
where $\alpha^i(\bbb{G},\cdot)$ denotes the composition of $\alpha(\bbb{G},\cdot)$ with itself $i$ times.
We call this induced graph the sunk graph generated by $v$.
The sunk graph generated by any vertex of each $\gamma(W(t))$, $t\ge 0$ has the following property.


\vspace{.05in}

\begin{lemma}
If $\bbb G$ is $(\beta,d\beta)$-resilient and the $n$-agent network is $(n-\kappa_{\beta,d\beta}(\bbb G))$-redundant, then for any time $t\ge 0$ and each vertex $v$ of $\gamma(W(t))$, there holds
$\bigcap_{i\in\scr{V}(v)} \scr P_i = 0$.
\label{easy}\end{lemma}

\vspace{.05in}

{\bf Proof of Lemma \ref{easy}:}
Since $\scr{V}(v)$ is the vertex set of the sunk graph generated by $v$,
it follows 
that $\scr{V}(v)$ contains all roots of $\gamma(W(t))$ whose number is at least $\kappa_{\beta,d\beta}(\bbb G)$. From Lemma \ref{lem:roots_uniqueness}, $\bigcap_{i\in\scr{V}(v)} \scr P_i = 0$. 
\hfill$\qed$

\vspace{.05in}


We also make use of the following concepts and results from \cite{le}.
Define a route over a given sequence of directed graphs $\mathbb{G}_1,\mathbb{G}_2,\ldots,\mathbb{G}_q$ with the same vertex set as
a sequence of vertices $i_0, i_1,\ldots, i_q$ such that
  $(i_{k-1},i_k)$  is an arc in $\mathbb{G}_k$ for all $k\in\{1,2,\ldots,q\}$.
  A route over a sequence of graphs which are all the same directed graph $\mathbb{G}$, is thus a directed   walk in $\mathbb{G}$.
  The definition implies that if $i_0, i_1,\ldots, i_q$ is a route over
$\mathbb{G}_1,\mathbb{G}_2,\ldots,\mathbb{G}_q$ and  $i_q,i_{q+1},\ldots,$ $i_p$ is a route over
$\mathbb{G}_q,\mathbb{G}_{q+1},\ldots,\mathbb{G}_p$,
then  the concatenated sequence $i_0, i_1,\ldots,i_{q-1}, i_q,i_{q+1},\ldots, i_p$
is a route over $\mathbb{G}_1,\mathbb{G}_2,\ldots,\mathbb{G}_{q-1},\mathbb{G}_{q}, \mathbb{G}_{q+1},\ldots,\mathbb{G}_p$.
This fact remains true if more than two sequences are concatenated.

More can be said if we focus exclusively on graphs with self-arcs. 
If $i=i_0, i_1,\ldots, i_q=j$ is a route over a sequence $\mathbb{G}_1,\mathbb{G}_2,\ldots,\mathbb{G}_q$,
then $(i,j)$ must be an arc in the composed graph $\mathbb{G}_q\circ\mathbb{G}_{q-1}\circ\cdots \circ\mathbb{G}_1$.
The converse is also true, namely that  if $(i,j)$ is an arc in $\mathbb{G}_q\circ\mathbb{G}_{q-1}\circ\cdots \circ\mathbb{G}_1$,
then there must exist vertices $i_1,\ldots, i_{q-1}$ for which $i=i_0, i_1,\ldots, i_q=j$ is a route over
$\mathbb{G}_1,\mathbb{G}_2,\ldots,\mathbb{G}_q$.
Moreover, if $\bbb{G}_{\tau_1},\bbb{G}_{\tau_2},\ldots,\bbb{G}_{\tau_p}$ is a subsequence of
$\mathbb{G}_1,\mathbb{G}_2,\ldots,\mathbb{G}_q$ with $p\le q$ and $i_0, i_1,\ldots, i_p$ being a route over
$\bbb{G}_{\tau_1},\bbb{G}_{\tau_2},\ldots,\bbb{G}_{\tau_p}$, then there must exist a route over $\mathbb{G}_1,\mathbb{G}_2,\ldots,\mathbb{G}_q$ which
contains $i_0, i_1,\ldots, i_p$ as a subsequence.
An important relation between routes and matrix multiplication is as follows. 

\vspace{.05in}

\begin{lemma} Let $S_1,S_2,\ldots S_q $ be a sequence of $m\times m$  stochastic matrices with positive diagonal entries whose graphs are respectively  
$\mathbb{G}_1,\mathbb{G}_2,\ldots,\mathbb{G}_q$. If  $j=i_0, i_1,\ldots, i_q=i$
 is a
route over $\mathbb{G}_1,\mathbb{G}_2,\ldots,\mathbb{G}_q$, then the   matrix product $P_{i_q}\cdots P_{i_{2}}P_{i_1}$
is a component of the   $ij$th block entry of
$P(S_q\otimes I)\cdots P(S_{2}\otimes I)  P(S_1\otimes I)$. \label{gum}\end{lemma}

\vspace{.05in}

Lemma \ref{gum} can be proved using the same argument as in the proof of Lemma 4 in \cite{le}.

It is easy to see from \eqref{eq:normalxstate_ls} that analysis of the algorithm under study involves the matrix product $\cdots P(W(t)\otimes I)\cdots  P(W(1)\otimes I)P(W(0)\otimes I)$ in which each $W(t)$, $t\ge 0$ is a $|\scr H|\times |\scr H|$ stochastic matrix with positive diagonal entries. Such a matrix product is a $d|\scr H|\times d|\scr H|$ block matrix whose each block is a projection matrix polynomial of the form 
$$\mu(P_1,P_2,\ldots,P_{|\scr H|}) = \sum_{i=1}^{b}\lambda_i P_{h_i(1)}P_{h_i(2)}\cdots P_{h_i(q_i)},$$
where $q_i$ and $b$ are positive integers, $\lambda_i $ is a real positive
   number, and each 
 $h_i(j)$, $j\in\{1,2,\ldots,q_i\}$  is an integer in $\{1,2,\ldots,|\scr H|\}$.

To study these block matrices, we need the following ``mixed matrix norm'' introduced in \cite{le}.
Let    $\|\cdot \|_{\infty}$ denote   the
  induced infinity  norm   and write
 $\R^{dm\times dm}$ for  the vector space of all $m\times m$  block matrices $Q = [Q_{ij}]$
whose $ij$th block entry     is a matrix $Q_{ij}\in\R^{d\times d}$.
  The mixed matrix norm of $Q\in\R^{dm\times dm}$, written $\|Q\|$, is defined as 
$\|Q\| = \|\langle Q\rangle \|_{\infty}$,
 where $\langle Q\rangle $ is the matrix in $\R^{m\times m}$  whose $ij$th entry is $\|Q_{ij}\|_2$.
It has been shown in \cite[Lemma 3]{le} that the mixed matrix norm  is 
sub-multiplicative. More can be said.

Let us agree to say that a  projection matrix polynomial
 $\mu(P_1,P_2,\ldots ,P_{|\scr H|})$ is    complete  if it has a    component
   $P_{h_i(1)}P_{h_i(2)}\cdots P_{h_i(q_i)}$ such that $\bigcap_{k=1}^{q_i}\scr P_{h_i(k)}=0$.
Such a complete component has the property that $\|P_{h_i(1)}P_{h_i(2)}\cdots P_{h_i(q_i)}\|_2<1$ \cite[Lemma 2]{le}. This property leads to a contraction condition for block matrices in $\R^{d|\scr H|\times d|\scr H|}$ whose blocks are projection matrix polynomials.

\vspace{.05in}

\begin{lemma}\label{lem:contraction}
    Let $S_1,S_2,\ldots S_q $ be a finite sequence of $|\scr H|\times |\scr H|$  stochastic matrices with positive diagonal entries and     
$M=P(S_q\otimes I)\cdots P(S_{2}\otimes I)  P(S_1\otimes I)$. If at least one entry in each block
row of $M$ is complete, then $M$ is a contraction in the mixed
matrix norm, i.e., $\|M\|<1$. 
\end{lemma}

\vspace{.05in}

The lemma is a direct consequence of Proposition 1 in~\cite{le}. 

\vspace{.05in}

\begin{proposition}
Suppose that \rep{assmp} holds. 
If $\bbb G$ is $(\beta,d\beta)$-resilient and the $n$-agent network is $(n-\kappa_{\beta,d\beta}(\bbb G))$-redundant, then there is a finite positive integer $\tau$ such that
for any  $p\geq \tau$ and $t\ge 0$,  the matrix
$P(W(t+p)\otimes I)\cdots  P(W(t+1)\otimes I)P(W(t)\otimes I)$
  is a contraction in the mixed  matrix norm.
\label{p}\end{proposition}

\vspace{.05in}


{\bf Proof of Proposition \ref{p}:}
Let $v$ be any vertex in $\scr{H}=\{1,2,\ldots,|\scr H|\}$ and $\scr{V}_t(v) = \{v\}\cup \alpha(\gamma(W(t)),v)\cup\cdots\cup\alpha^{|\scr H|-1}(\gamma(W(t)),v)$
be the vertex set of $\gamma(W(t))$'s
sunk graph generated by $v$. From Lemma \ref{easy}, $\bigcap_{i\in\scr{V}_t(v)} \scr P_i = 0$ for all $t\ge 0$. Since $\scr{V}_t(v)\subset \scr{H}$ for all $t$,
the total number of possibly distinct $\scr{V}_t(v)$ is finite.
We use $h(v)$ to denote this finite number.
Let $b_t$ denote the cardinality of $\scr{V}_t(v)$ and let $\bar b$ be the maximum of all possible $b_t$, i.e., $\bar b = \max_t b_t$. It is clear that $\bar b\leq |\scr H|$.
Set $r=\frac{1}{2}(\bar b-1)\bar b$ and $\tau(v)=(r-1)h(v)$. Then, the pigeonhole principle (or Dirichlet's box principle) guarantees that at least  one vertex set in the sequence $\scr{V}_t(v),\scr{V}_{t+1}(v),\ldots,\scr{V}_{t+\tau(v)}(v)$ appears at least $r$ times for any $t\ge 0$.
We use $\scr{U}$ to denote this vertex set and let $b$ be its cardinality.
Let $\gamma(W(\tau_1)),\gamma(W(\tau_2)),\ldots,\gamma(W(\tau_q))$ be the subsequence of
$\gamma(W(t)),\gamma(W(t+1)),\ldots,\gamma(W(t+\tau(v)))$
which includes all those graphs in the sequence whose vertex set of the sunk graph generated by $v$
is $\scr{U}$.
It is easy to see that $\tau(v)\ge q\geq r \geq \frac{1}{2}(b-1)b$.
Since the $n$-agent network is $(n-\kappa_{\beta,d\beta}(\bbb G))$-redundant and each $\gamma(W(t))$, $t\ge 0$ is rooted (cf. Lemma \ref{lem:root}), $b\ge \kappa_{\beta,d\beta}(\bbb G)>1$.

We next show that there exists at least one complete
block entry in the $v$th row of
$M=P(W(t+p)\otimes I)\cdots  P(W(t+1)\otimes I)P(W(t)\otimes I)$ for any $p\ge \tau(v)$.


To simplify notation, let $\bbb{H}_1=\gamma(W(\tau_q)), \bbb{H}_2=\gamma(W(\tau_{q-1})), \ldots, \bbb{H}_q=\gamma(W(\tau_1))$.
Set $\Sigma_k= 1+2+\cdots + k = \frac{1}{2}k(k+1)$ for each $k\in\{1,2,\ldots,b-1\}$.
Partition the sequence $\bbb{H}_1, \bbb{H}_2,\ldots,\bbb{H}_q$ into
$b-1$ successive subsequences:
$\scr{H}_1= \bbb{H}_1$,
$\scr{H}_2= \bbb{H}_2,\bbb{H}_3$,
$\scr{H}_3= \bbb{H}_4,\bbb{H}_5,\bbb{H}_6$,
$\ldots,$
$\scr{H}_{b-2} = \bbb{H}_{\Sigma_{b-3}+1},\bbb{H}_{\Sigma_{b-3}+2},\ldots,\bbb{H}_{\Sigma_{b-2}}$,
and $\scr{H}_{b-1} = \bbb{H}_{\Sigma_{b-2}+1},\bbb{H}_{\Sigma_{b-2}+2},\ldots,\bbb{H}_q$.
Note that each $\scr{H}_i$ is of length $i$ except for the last which must be of length
$q-\Sigma_{b-2}\geq b-1$.
Let $\scr{U}=\{i_1,i_2,\ldots,i_b\}$ and $i_1=v$.
Since $\scr{U}$ is the vertex set of the sunk graph generated by $v$ for each $\bbb{H}_i$
and $b>1$, vertex $i_1$ must have a neighbor $i_2$ in $\bbb{H}_1$. Then, there exists a route over $\scr{H}_1$ from $i_2$ to $i_1$.
Since $b\le |\scr H|$, from Lemma~\ref{kobs}, vertex $i_2$ has at least three neighbors
in $\bbb{H}_{2}\circ \bbb{H}_3$, implying that vertex $i_2$ must have a neighbor $i_3$ in $\bbb{H}_{2}\circ \bbb{H}_3$
which is not in the set $\{i_1,i_2\}$.
Thus, there exists a route over $\scr{H}_2$ from $i_3$ to $i_2$.
Repeating this argument for all $\scr{H}_i$ will yield a sequence of distinct vertices $i_1,i_2,\ldots,i_b$ such that
there exists a route over $\scr{H}_k$ from $i_{k+1}$ to $i_{k}$ for all $k\in\{1,2,\ldots,b-1\}$.
Let
$i_{k+1}=j_{\Sigma_{k}},j_{\Sigma_{k}-1},\ldots,j_{\Sigma_{k-1}}=i_k$
denote the route from $i_{k+1}$ to $i_{k}$ 
over $\scr{H}_k = \bbb{H}_{\Sigma_{k-1}+1},\ldots,\bbb{H}_{\Sigma_{k}}$ for each
$k\in\{1,2,\ldots,b-2\}$, 
and let $i_{b}=j_{q},j_{q-1},\ldots,j_{\Sigma_{b-2}}=i_{b-1}$
denote the route from $i_{b}$ to $i_{b-1}$ over $\scr{H}_{b-1} = \bbb{H}_{\Sigma_{b-2}+1},\ldots,\bbb{H}_q$. 
Then, $j_q,j_{q-1},\ldots,j_0$ must be a route
over the overall sequence $\bbb{H}_q,\bbb{H}_{q-1},\ldots,\bbb{H}_1$. In particular,
$j_{\Sigma_{k-1}}=i_k$ for all $k\in\{1,2,\ldots,b-1\}$ and  $j_q=i_b$.
It follows that
$i_b=j_q,j_{q-1},\ldots,j_0 = i_1=v$  is a route over $\gamma(W(\tau_1)),\gamma(W(\tau_2)),\ldots,\gamma(W(\tau_q))$.
Since $\gamma(W(\tau_1)),\gamma(W(\tau_2)),\ldots,\gamma(W(\tau_q))$ is a subsequence
of $\gamma(W(t)),\gamma(W(t+1)),\ldots,\gamma(W(t+p))$ and
all $\gamma(W(t))$, $t\ge 0$ have self-arcs at all vertices,
there must exist a route $k_0,k_1,\ldots,k_s=v$ over $\gamma(W(t)),\gamma(W(t+1)),\ldots,\gamma(W(t+p))$
which contains $i_b=j_q,j_{q-1},\ldots,j_0 = i_1 =v$ as a subsequence.
From Lemma \ref{gum}, the matrix product $P_{k_{s}}\cdots P_{k_{2}}P_{k_{1}}$
must be a component of the $(vk_0)$th block entry of $M$.
Recall the definition of $\scr{U}$ which implies that $\bigcap_{i\in\scr{U}} \scr P_i = 0$. Since each element of $\scr{U}$ appears in $k_1,k_{2},\ldots,k_s$  at least once,
the $(vk_0)$th block entry of $M$ is complete.

Note that the preceding arguments apply to all  $v\in\scr H$. It follows that 
at least one entry in each row of $M$ is a complete projection matrix polynomial provided $p\ge \tau(v)$ for all $v\in\scr H$.
From Lemma \ref{lem:contraction}, $M$ is a contraction in the mixed matrix norm.
\hfill$\qed$


\vspace{.05in}

The following theorem establishes the correctness of the algorithm for the unique least squares solution case. 

\vspace{.05in}

\begin{theorem}
Suppose that $Ax=b$ has a unique least squares solution. 
If $\bbb G$ is $(\beta,d\beta)$-resilient and the $n$-agent network is $(n-\kappa_{\beta,d\beta}(\bbb G))$-redundant, then there exists a nonnegative constant $\lambda<1$
for which all $x_i(t)$, $i\in\scr H$ converge to the least squares solution as $t\rightarrow\infty$
as fast as $\lambda^t$ converges to zero.
\label{main1}\end{theorem}


\vspace{.05in}

{\bf Proof of Theorem \ref{main1}:}
From Proposition \ref{p}, there is a finite positive integer $\tau$ such that
for any  $p\geq \tau+1$,  the matrix
$P(W(t+p)\otimes I)\cdots  P(W(t+1)\otimes I)P(W(t)\otimes I)$
  is a contraction in the mixed  matrix norm for all $t\ge 0$.
Write $\scr W_\tau$ for the set of all distinct subsequences
 $W(k\tau),W(k\tau+1),\ldots, W((k+1)\tau)$, $k\geq 0$
 encountered along any trajectory of \eqref{eq:normalxstate_ls}. From Lemma \ref{lem:root}, $\scr W_\tau$ is a compact set, implying that 
\begin{align*}
    \lambda &= \Big(\sup_{k\in\{0,1,2,\ldots\}} \big\|P(W((k+1)\tau)\otimes I)\cdots \\
  & \;\;\;\;\;\;\;\;\;\;\;\;\;\;\;\;\;\; P(W(k\tau+1)\otimes I)P(W(k\tau)\otimes I)\big\|\Big)^{\frac{1}{\tau}} <1.
\end{align*}
The statement of the theorem then follows from this observation and the fact that $\|\cdot\|$ is sub-multiplicative. 
\hfill$\qed$

\vspace{.05in}

We next consider the case when $Ax = b$ has multiple least squares solutions, using the subspace ``quotient out'' technique from \cite{le}, and begin with 
the following lemma.

\vspace{.05in}

\begin{lemma}
Let $Q'$ be any matrix whose columns form an orthonormal basis for the orthogonal complement of $\bigcap_{i\in\scr H}\scr{P}_i$ and define $\bar{P}_i = QP_iQ'$ for each $i\in\scr H$.
Then, 
\begin{itemize}
    \item[1)] Each $\bar{P}_i$, $i\in\scr H$ is an orthogonal projection matrix;
    \item[2)] Each $\bar{P}_i$, $i\in\scr H$  satisfies $QP_i = \bar{P}_i Q$;
    \item[3)] For any nonempty subset $\scr{E}\subset\scr H$, there holds $\bigcap_{i\in\scr{E}} \bar{\scr{P}}_i =0$ if and only if $\bigcap_{i\in\scr{E}}\scr P_i=\bigcap_{i\in\scr{H}}\scr P_i$.
\end{itemize}
\label{vlad}\end{lemma}

\vspace{.05in}

The lemma is a direct consequence of Lemma 7 in \cite{le_automatica}. 

\vspace{.05in}

\begin{lemma}
If $\bbb G$ is $(\beta,d\beta)$-resilient and the $n$-agent network is $(n-\kappa_{\beta,d\beta}(\bbb G))$-redundant, then for any time $t\ge 0$ and each vertex $v$ of $\gamma(W(t))$, there holds
$\bigcap_{i\in\scr{V}(v)} \bar{\scr P}_i = 0$.
\label{noteasy}\end{lemma}

\vspace{.05in}

{\bf Proof of Lemma \ref{noteasy}:}
Since $\scr{V}(v)$ is the vertex set of the sunk graph generated by $v$, it follows that $\scr{V}(v)$ contains all roots of $\gamma(W(t))$ whose number is at least $\kappa_{\beta,d\beta}(\bbb G)$. 
From Corollary \ref{coro:global_least_squares}, 
$\argmin_{x}\sum_{i\in\scr{V}(v)}
    \|A_ix-b_i\|_2^2=\scr X^*$.
This fact and Proposition \ref{prop:common_least_squares} imply that 
$\bigcap_{i\in\scr{V}(v)}\scr P_i=\bigcap_{i\in\scr{H}}\scr P_i$. 
From property 3) of Lemma \ref{vlad}, $\bigcap_{i\in\scr{V}(v)} \bar{\scr P}_i = 0$. 
\hfill$\qed$


\vspace{.05in}

\begin{theorem}
Suppose that $Ax=b$ has more than one least squares solution. If $\bbb G$ is $(\beta,d\beta)$-resilient and the $n$-agent network is $(n-\kappa_{\beta,d\beta}(\bbb G))$-redundant, then there exists a nonnegative constant $\lambda<1$
for which all $x_i(t)$, $i\in\scr H$ converge to the same least squares solution as $t\rightarrow\infty$
as fast as $\lambda^t$ converges to zero.
\label{main2}\end{theorem}


{\bf Proof of Theorem \ref{main2}:}
Property 2) of Lemma \ref{vlad} implies that $QP_iP_j = \bar{P}_i\bar{P}_jQ$ for all
$i,j\in\scr H$.  Define $\bar{y}_i = Qy_i$ for each $i\in\scr H$. Then, from \eqref{eq:y_i_update}, for all $i\in\scr H$,
\begin{align}
    \bar y_i(t+1) 
    = \bar P_i\Big(w_{ii}(t)\bar y_i(t) + \sum_{k\in\scr N_i\cap\scr H}w_{ik}(t)\bar y_k(t)\Big). \label{nna1}
\end{align}
Define $z_i = y_i-Q'\bar{y}_i$ for each $i\in\scr H$.
Since $Qz_i = Qy_i - \bar{y}_i$, $Qz_i = 0$ for all $i\in\scr H$.  Then, $z_i(t)\in\bigcap_{j\in\scr H} \scr{P}_j$ for all $i\in\scr H$, which implies that $P_jz_i(t) = z_i(t)$ for any $i,j\in\scr H$.
With these facts and property 2) of Lemma \ref{vlad}, \rep{nna1} leads to 
\eq{z_i(t+1) =
w_{ii}(t)z_i(t) + \sum_{k\in\scr N_i\cap\scr H}w_{ik}(t)z_k(t),\label{dinner}}
a standard discrete-time linear  consensus process \cite{reachingp1}.
Note that update \rep{nna1} has exactly the same form as \eqref{eq:y_i_update} except that $P_i$ is replaced by $\bar{P}_i$.
From property 1) of Lemma \ref{vlad},  all $\bar{P}_i$, $i\in\scr H$ are also orthogonal  projection matrices. 
Property 3) of Lemma \ref{vlad} implies that $\bigcap_{i\in\scr{H}} \bar{\scr{P}}_i =0$.
Moreover, Lemma \ref{noteasy} guarantees the same property of $\bar{\scr P}_i$ as that of $\scr P_i$ in Lemma~\ref{easy}. With these facts, Theorem \ref{main1}
  is applicable to the system of iterations \rep{nna1}. Thus, all $\bar y_i(t)$, $i\in\scr H$ defined by  \rep{nna1} converge to zero exponentially fast. 
Since the graph of each $W(t)$, $t\ge 0$ is rooted (cf. Lemma \ref{lem:root}),  all $z_i(t)$, $i\in\scr H$ defined by  \rep{dinner} converge to the same limit $z^*$ exponentially fast \cite[Theorem 2]{reachingp1}. From the preceding, $z^*\in\bigcap_{i\in\scr H}\scr{P}_i$. 
Therefore, all $x_i(t)$, $i\in\scr H$ defined by  \eqref{eq:x_ls_ana} converge to the same limit exponentially fast, and the limit 
is $x^* + z^*$ with $x^*\in\scr X^*$ which is a least squares solution to $Ax=b$.
\hfill$\qed$


\section{Conclusion}

This paper has proposed a distributed least squares algorithm for solving a system of linear algebraic equations over a fixed multi-agent network, which converges exponentially fast and achieves full resilience in the presence of Byzantine agents provided appropriate redundancy in both graph connectivity and objective functions is established. The proposed algorithm and its convergence results can be easily extended to non-stationary networks provided that the time-varying neighbor graphs are always $(\beta,d\beta)$-resilient. 
Since the algorithm borrows the same design ideas from a recent resilient distributed optimization algorithm \cite{acc2023}, it ``inherits'' the same limitations from the algorithm there (see discussions in Section 5 of \cite{acc2023}). 
A particular limitation is that it can hardly cope with  high-dimensional cases. 

An important future direction is thus to tackle the challenging high-dimensional issue. Although it is a natural idea to appeal to  communication-efficient schemes in which each agent only needs to transmit low-dimensional signals (e.g., entry- or block-wise updating \cite{cdc20,block}), our limited simulations indicate that simply partitioning high-dimensional state vectors into low-dimensional blocks in communication and computation is not promising, if without any additional design. Other possible approaches include exploiting consensus fusion with reduced information \cite{cdc22},  leveraging dimension-independent filtering \cite{gupta2021byzantine,acc22}, and combining these techniques together \cite{toread}. 



\bibliographystyle{unsrt}
\bibliography{push,resilience}

\end{document}